\documentclass{iaus}
\usepackage{graphicx}

\title[The peculiar Be phenomenon] 
{The B[e] phenomenon in the Milky Way and Magellanic Clouds}

\author[Anatoly Miroshnichenko et al.]   
{Anatoly Miroshnichenko$^1$, Nadine Manset$^2$, Francesco
Polcaro$^3$, Corinne Rossi$^4$, Sergey Zharikov$^5$}

\affiliation{$^1$Dept. of Physics \& Astronomy, University of North Carolina at Greensboro, \\
P.O. Box 26170, Greensboro, NC 27402--6170, USA  \\email: {\tt
a$\_$mirosh@uncg.edu}\\
$^2$CFHT Corporation, 65--1238 Mamalahoa Highway, Kamuela, HI 96743\\
$^3$ Istituto di Astrofisica Spaziale e Fisica Cosmica, INAF, Via del Fosso del Cavaliere 100, 00133, Roma, Italy\\
$^4$ Universit\'a La Sapienza Roma - Pza A Moro 5, I--00162 Roma, Italy\\
$^5$ Instituto de Astronom\' ia, Universidad Nacional Aut\'onoma de
M\'exico, Apartado Postal 877, 22800, Ensenada, BC, Mexico }

\pubyear{2010}
\volume{272}  
\pagerange{1--2}
 \jname{Active OB stars: structure, evolution,
mass loss and critical limits} \editors{C. Neiner, G. Wade, G.
Meynet \& G. Peters, eds.}

\begin{document}

\maketitle

\begin{abstract}
Discovered over 30 years ago, the B[e] phenomenon has not yet
revealed all its puzzles. New objects that exhibit it are being
discovered in the Milky Way, and properties of known objects are
being constrained. We review recent findings about objects of this
class and their subgroups as well as discuss new results from
studies of the objects with yet unknown nature. In the Magellanic
Clouds, the population of such objects has been restricted to
supergiants. We present new candidates with apparently lower
luminosities found in the LMC. \keywords{stars: early-type;
infrared: stars; (stars:) circumstellar matter}
\end{abstract}

\firstsection 
\section{Introduction}

The B[e] phenomenon was observationally defined by \cite[Allen \&
Swings (1976)]{AllenSwings76} on the basis of optical spectroscopic
and near-infrared photometric data. It refers to the simultaneous
presence of forbidden line emission (e.g., [O {\sc I}], [Fe {\sc
II}], [N {\sc II}], and sometimes [O {\sc III}] lines) in addition
to permitted line emission (e.g., Balmer and Fe {\sc II} lines) and
large IR excesses in the spectra of B-type stars. These
observational features make it different from the Be phenomenon. The
presence of forbidden lines indicates that the gaseous component of
the circumstellar (CS) envelopes is more extended than in Be stars.
The large IR excess is a manifestation of CS dust which is not
present in Be stars. Also, the B[e] phenomenon occurs in a wider
variety of objects than the Be phenomenon (see \cite[Miroshnichenko
2006]{Miroshnichenko06} for a recent review).

Allen \& Swings have already noticed the variety of objects with the
B[e] phenomenon. Some twenty years later \cite[Lamers et al.
(1998)]{Lamers98} summarized available data and concluded that the
B[e] phenomenon occurs in four stellar groups with well-understood
nature and evolutionary status (Herbig Ae/Be stars, symbiotic
binaries, supergiants, and compact Planetary Nebulae). At the same
time, nearly half of the original list of 65 objects with the B[e]
phenomenon remained unclassified. A few of them were well-studied
(e.g., FS\,CMa = HD\,45677 and V742\,Mon = HD\,50138), but their
derived properties did not allow to fit them within any of the above
groups. Other unclassified objects have not been studied enough to
classify them until recently.

A few years ago \cite[Miroshnichenko (2007)]{Miroshnichenko07} and
\cite[Miroshnichenko et al. (2007)]{Miroshnichenko_etal07} analyzed
both historic and their own data on the unclassified objects and
concluded that most of them can be separated into a new group. The
group was called FS\,CMa type objects after the prototype object
with the B[e] phenomenon (\cite[Swings 2006]{Swings06}). These
authors also showed that FS\,CMa objects are neither
pre-main-sequence Herbig Ae/Be stars nor symbiotic binaries, their
luminosity range ($\log$ L/L$_{\odot} \sim$ 2.5--4.5) is below that
of supergiants, and they are unlikely to be at the post-AGB
evolutionary stage. The main observational features of this group
are described in the other paper by Miroshnichenko et al. in these
proceedings. One of these features is a very strong emission-line
spectrum which is hard to explain by the evolutionary mass loss from
a single star of the described luminosity range (\cite[Vink, de
Koter, \& Lamers 2001]{Vink01}). Therefore, it was suggested that
FS\,CMa objects are binary systems observed after a rapid
mass-exchange phase (\cite[Miroshnichenko 2007]{Miroshnichenko07}).
No direct mass transfer seems to be currently observed in any of
these objects.

Currently, only a few objects with the B[e] phenomenon can be called
unclassified due to insufficient data. All Herbig Ae/Be stars, not
only seven from the original list, exhibit the B[e] phenomenon.
Therefore, the original list of 65 objects has been expanded
significantly. Among the variety of objects with the B[e]
phenomenon, only supergiants and FS\,CMa type objects seem to form
dust while having a B-type star in their content. Here we will
concentrate on the expansion of the list of objects which belong to
the FS\,CMa group in the Milky Way and the Large Magellanic Cloud
(LMC).

\section{Finding new candidates.}

Another feature of the FS\,CMa group is a sharp decrease of the IR
flux at $\lambda > 10 \mu$m. It allowed \cite[Miroshnichenko et al.
(2007)]{Miroshnichenko_etal07} to expand the group by finding nine
IRAS sources (whose fluxes were accurately measured in three IRAS
photometric bands at 12, 25, and 60 $\mu$m) with such a flux
behavior that positionally coincide with early-type emission-line
stars. However, this procedure only works for relatively bright IR
sources, because the sensitivity of IRAS decreases with wavelength.
Also, the lowest-mass post-AGB objects, RV\,Tau type stars, turned
out to have IRAS colors within the same range.

Nevertheless, a new photometric criterion for FS\,CMa objects was
established in the same paper. It was found that the observed $J-K$
color-indices of FS\,CMa objects exceed $\sim$1.3 mag, while RV\,Tau
stars show bluer colors. We decided to use this criterion in
combination with not very large optical colors and search for new
candidates with the B[e] phenomenon in a catalog of emission-line
stars by \cite[Kohoutek \& Wehmeyer (1997)]{KW97}, which we
cross-correlated with the NOMAD catalog by \cite[Zacharias et al.
(2005)]{Zacharias05}.

\begin{figure}[htb]
\vspace*{-1.0 cm}
\begin{center}
\begin{tabular}{ll}
\resizebox{7.0cm}{!}{\includegraphics{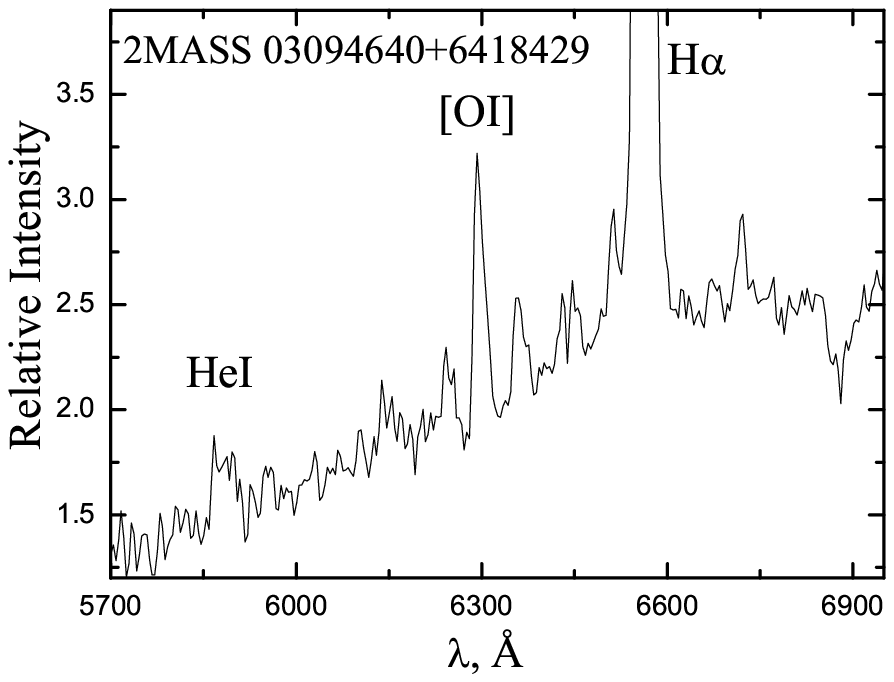}}
&
\resizebox{7.0cm}{!}{\includegraphics{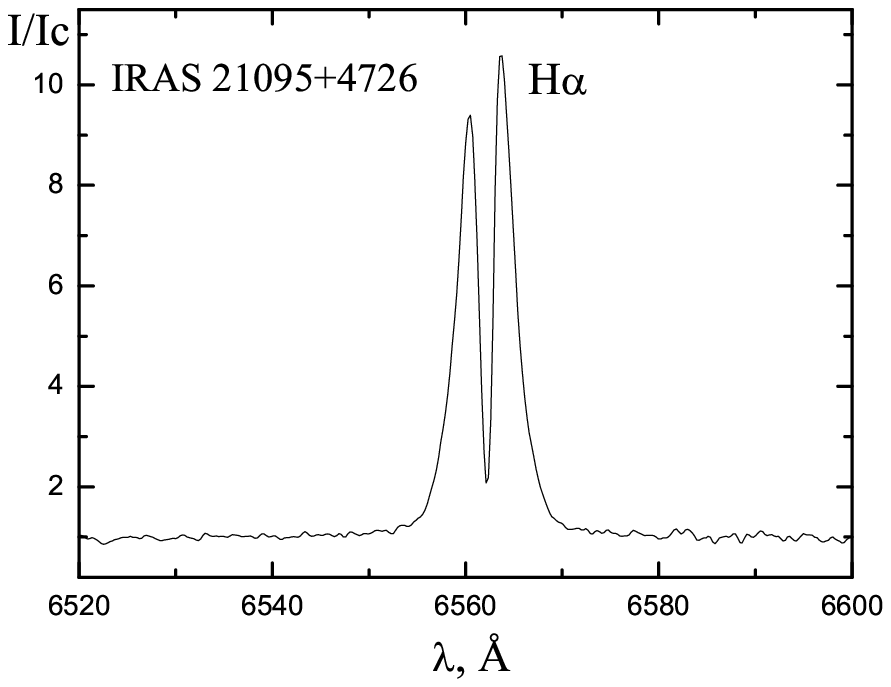}}\\
\end{tabular}
\vspace*{-0.5 cm} \caption{Left panel: Low-resolution spectrum of
2MASS\,03094640+6418429 obtained at the 1.82-m telescope of the
Padova Observatory ($R \sim$ 1500). Right panel: High-resolution
spectrum of IRAS\,21095+4726 obtained at the 2.12-m telescope of the
San Pedro Martir Observatory ($R \sim$ 15000).}
   \label{talk_Miroshnichenko_fig1}
\end{center}
\end{figure}

As a result, we found sixteen new candidates. Five of them, randomly
picked, were observed and the presence of forbidden lines was found
in all of them. Additionally, a very faint object with the B[e]
phenomenon ($V \sim$ 17 mag) was accidentally found. It turned out
to positionally coincide with the 2MASS source 03094640+6418429.
Fragments of our spectra of some of these objects are shown in
Fig.\,\ref{talk_Miroshnichenko_fig1}. They are most likely FS\,CMa
objects, because luminosity sensitive absorption lines (e.g., Si
{\sc III} 5739 \AA, \cite[Miroshnichenko et al.
2004]{Miroshnichenko04}) typical for supergiants, were not found in
their spectra. Other newly discovered and spectroscopically
confirmed objects with the B[e] phenomenon include IRAS\,02110+6212
= VES\,723, IRAS\,21263+4927, IRAS\,20090+3809, and MWC\,485.

We continue searching the NOMAD catalog and have already found
another dozen of new candidates. The photometric search is not
exhaustive, because we only look for objects with not very different
magnitudes in the blue and red band. Redder objects may be just very
cool stars. Nevertheless, currently the number of FS\,CMa objects
and candidates approaches 70.

\section{New candidates in the Large Magellanic Cloud.}

The first object (R\,126) with the B[e] phenomenon in the Magellanic
Clouds was reported by \cite[Zickgraf et al. (1985)]{Zickgraf85}.
Another seven were added by \cite[Zickgraf et al.
(1986)]{Zickgraf86} and four more by \cite[Gummersbach, Zickgraf, \&
Wolf (1995)]{Gummersbach95}. All these objects are located beyond
the main-sequence. Ten of them are supergiants with luminosities
$\log$ L/L$_{\odot} = 4.7-6.1$, while the remaining two (Hen\,S59
and Hen\,S137) are late B-type stars with a luminosity of $\log$
L/L$_{\odot} \sim 4$.

It is easier to separate supergiants from less luminous stars in the
Clouds with their low interstellar extinction than in the Milky Way.
All the supergiants with the B[e] phenomenon in the LMC have optical
brightnesses of $V = 11-13$ mag objects, while Hen\,S59 and
Hen\,S137 have $V \sim$ 14 mag. In an attempt to find more
candidates to the list of objects with the B[e] phenomenon in the
LMC, we positionally cross-correlated a catalog of optical
photometry by \cite[Zaritsky et al. (2004)]{Zaritsky04} and the
2MASS catalog (\cite[Cutri et al. 2003]{Cutri03}).

\begin{table}[htb]
  \begin{center}
  \caption{Candidates for objects with the B[e] phenomenon in LMC}
  \label{tab1}
 {\scriptsize
  \begin{tabular}{|l|c|c|c|c|c|} \hline
{\bf Name}   & {\bf R.A.} & {\bf Dec.}    & $V$   & $K$   & $J-K$ \\
\hline
ARDB 54      & 4:54:43.5  & $-$70:21:27.8 & 12.77 & 11.55 & 1.15  \\
0218-0100858 & 5:45:29.5  & $-$68:11:45.7 & 14.02 & 11.48 & 1.62  \\
0203-0138943 & 5:41:43.7  & $-$69:37:38.3 & 14.11 & 11.14 & 2.00  \\
0181-0125572 & 5:27:47.6  & $-$71:48:52.6 & 14.24 & 11.77 & 1.72  \\
BE74 540     & 5:12:09.1  & $-$71:06:49.7 & 14.27 & 12.11 & 1.61  \\
BE74 580     & 5:24:17.4  & $-$71:31:50.0 & 14.56 & 12.24 & 1.55  \\
0225-0105286 & 5:24:57.9  & $-$67:24:57.9 & 14.67 & 12.51 & 1.40  \\
LHA120-N 148B& 5:31:42.2  & $-$68:34:53.9 & 15.42 & 13.41 & 1.50  \\
AL 190       & 5:26:30.7  & $-$67:40:36.5 & 15.69 & 12.48 & 2.12  \\
\hline
 \end{tabular}
  }
 \end{center}
\vspace{1mm}
 \scriptsize{
 {\it Comments:}\\
The objects' names are from \cite[Bohanan \& Epps (1974)]{BE74} –-
BE74, \cite[Henize (1956)]{Henize56} –- LHA120, \cite[Andrews \&
Lindsay (1964)]{AL64} –- AL, \cite[Ardeberg et al.
(1972)]{Ardeberg72} –- ARDB, and the rest is from the USNO--B1.0
survey (\cite[Monet et al. 2003]{Monet03}).\\
The coordinates are given for the epoch 2000.0. }
\end{table}

\begin{figure}[htb]
\begin{center}
\begin{tabular}{ll}
\resizebox{2.9in}{!}{\includegraphics{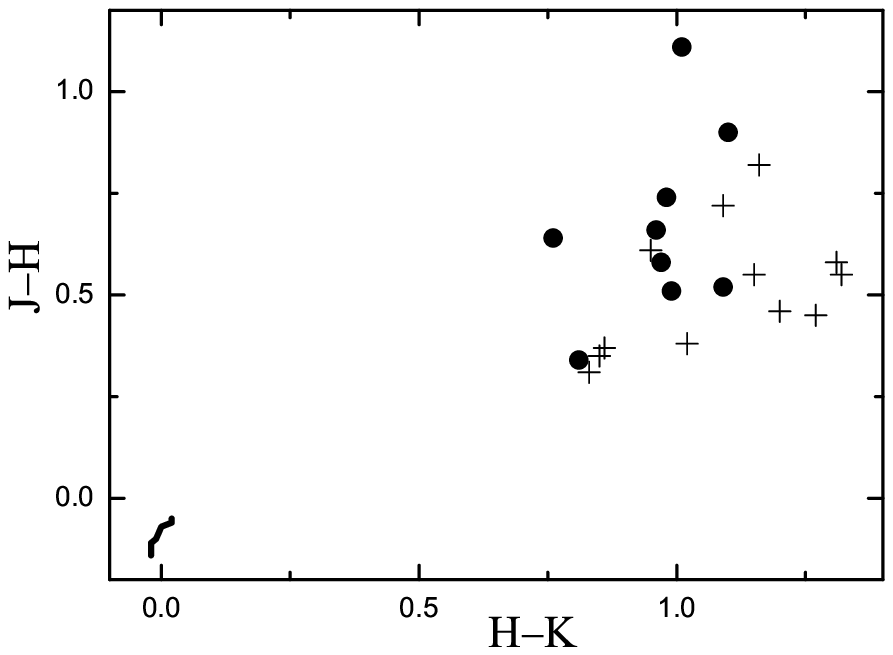}}
&
\hspace*{-1.0cm} \resizebox{2.8in}{!}{\includegraphics{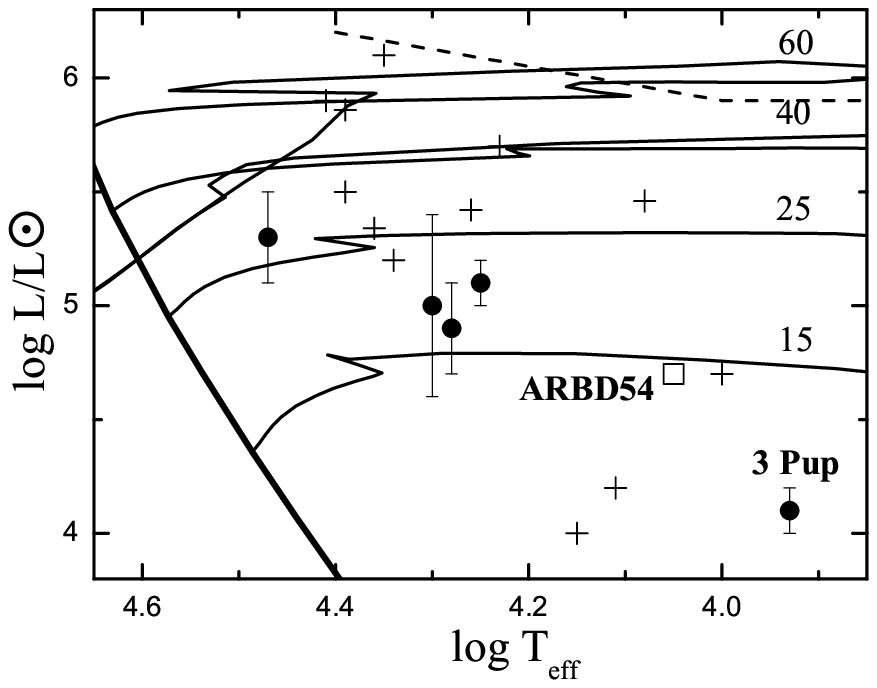}}\\
\end{tabular}
\caption{Left panel: A $(J-H) \sim (H-K)$ diagram for the known
Clouds objects with the B[e] phenomenon (pluses) and the objects
from Table\,\ref{tab1} (circles). The solid line in the lower left
corner represents intrinsic colors of B-type supergiants
(\cite[Wegner 1994]{Wegner94}). Right panel: A Hertzsprung-Russell
diagram for supergiants with the B[e] phenomenon and
well-constrained luminosities in the Milky Way (circles) and in the
Clouds (pluses). The newly found candidate ARDB 54 is shown by an
open square. The solid lines are the zero-age main-sequence and
evolutionary tracks for single stars (\cite[Schaller et al.
1992]{Schaller92}) labeled with initial masses in solar units. The
dashed line is shows the Humphreys-Davidson stability limit.}
   \label{talk_Miroshnichenko_fig2}
\end{center}
\end{figure}

In total, we found nearly 100 positionally close objects that have
slightly reddened colors of B-type stars and large IR color-indices,
but for only nine of them the optical-IR position offset does not
exceed 1 arcminute. These objects are listed in Table\,\ref{tab1}.
One of these objects, ARDB\,54, is a supergiant (see right panel of
Fig.\,\ref{talk_Miroshnichenko_fig2}). All the others have $V =
14.0-15.7$ mag and represent lower luminosity objects.

A near-IR color-color diagram for the recognized objects with the
B[e] phenomenon in the Magellanic Clouds and the objects from
Table\,\ref{tab1} is shown in the left panel of
Fig.\,\ref{talk_Miroshnichenko_fig2}. If forbidden line emission is
found in their spectra, then they probably belong to the FS\,CMa
group. We plan to obtain both photometric and spectroscopic
observations of these objects in the nearest future.

\section{Conclusions.}

Using our photometric criteria for separation of objects with the
B[e] phenomenon, we found nearly 20 new candidates in the Milky Way
and eight in the LMC. One new supergiant candidate is also found in
LMC. The presence of the B[e] phenomenon in five Galactic objects
was confirmed. Thus, we expanded the Galactic FS\,CMa group to
nearly 50 members and 20 candidates. The number of objects with the
B[e] phenomenon in the LMC may be doubled, if its presence in the
spectra of the reported candidates is confirmed.\\

{\bf Acknowledgements}. This research has made use of the SIMBAD
database operated at CDS, Strasbourg, France, and of data products
from the Two Micron All Sky Survey, which is a joint project of the
University of Massachusetts and the Infrared Processing and Analysis
Center/California Institute of Technology, funded by the National
Aeronautics and Space Administration and the National Science
Foundation.

\end{document}